\documentclass[aps,prl,preprint,]{revtex4}

\newcommand{\beq}{\begin{equation}}
\newcommand{\eeq}{\end{equation}}
\newcommand{\bea}{\begin{eqnarray}}
\newcommand{\eea}{\end{eqnarray}}

\begin{document}

\title{A GENERALIZED SCHR\"ODINGER EQUATION FOR LOOP QUANTUM COSMOLOGY}

\author{D. C. SALISBURY and A. SCHMITZ}

\affiliation{DEPARTMENT OF PHYSICS, AUSTIN COLLEGE,
Sherman, TX  75090, USA\\E-mail:
dsalisbury@austincollege.edu}

\begin{abstract}
A temporally discrete Schroedinger time evolution equation is proposed for isotropic quantum cosmology coupled to a massless scalar source. The approach employs dynamically determined intrinsic time and produces the correct semiclassical limit.
\end{abstract}

\maketitle
\section{Introduction}
Popular approaches to loop quantum cosmology recover a notion of time within a "frozen time" formalism through requiring that the Hamiltonian constraint annihilate physical states.\cite{bojowaldH02} It is claimed that the resulting states encode unique correlations between dynamical observables and intrinsically defined time. In contrast we present here a simple isotropic cosmological model with a massless scalar source in which we argue that it is possible to formulate a unique quantum time evolution. Furthermore, we demonstrate explicitly that this evolution produces the correct semi-classical limit. The program utilizes intrinsically defined time, and is motivated by the recognition that classical cosmological variables expressed in terms of intrinsic time can be shown to be invariant  under the canonically realized group of time coordinate transformations.\cite{ps04,hss04}  This work is based on an improved understanding of the nature of the diffeomorphism-induced canonical symmetry group in which it is recognized that lapse functions must be retained as canonical variables.\cite{pss:1997pr} We will first discuss the classical implications of this group from four different but equivalent perspectives, and then we will propose a generalized intrinsic-time-dependent Schr\"odinger equation.

\section{Classical intrinsic time and canonical reparameterization invariance}
We will consider an isotropic cosmological model with expansion factor $a(t)\ell_P $, massless scalar source $\phi(t)\sqrt{m_P/t_P} $ and lapse function $ N(t)\ell_P$, where for later convenience we express all fields in Planckian units (so that $ a(t)$, $\phi(t)$, and $N(t)$, as well as the time coordinate $t$ are all dimensionless). The reduced Lagrangian takes the form $L = \hbar \left( -\frac{3 a \dot a^2}{4 \pi  N} + \frac{a^3 \dot \phi^2}{2 N} \right)$. The resulting Hamiltonian is $H = \frac{N}{\hbar} \left(-\frac{\pi p_a^2}{3 a} + \frac{ p_\phi^2}{2 a^3} \right) + \lambda p_N$ where $\lambda$ is an arbitrary positive-definite function of $t$, and the factors multiplying $N$ and $\lambda$ are primary constraints. The Lagrangian model is covariant under infinitesimal reparameterizations in time of the form $t' = t - N^{-1} \xi(t)$, and corresponding variations in the canonical variables are faithfully generated by the phase space generator  $G_\xi := \frac{\xi}{\hbar} \left( -\frac{\pi p_a^2}{3 a}  + \frac{ p_\phi^2}{2 a^3} \right) + \dot \xi p_N$. We shall choose as our intrinsic time $T$ the square of the expansion factor, thus in terms of the general solution of the equations of motion $T(t) = a^2(t) =   \left(N_0t + \int_0^t dt' \int_0^{t'} dt'' \lambda(t'') + a_0^3 \right)^{2/3}$. The naught subscript signifies variables evaluated at time $t = 0$. There are now four equivalent ways to construct reparameterization invariants:
\begin{itemize}
\item Perform the time reparameteriztion $T(t)$. Thus the invariant variables are $\phi(T) = \phi(t(T)) = \phi_0 \pm \sqrt{\frac{1}{6 \pi}}  \left( \frac{3}{2} \log T - 3 \log a_0 \right) =  \phi_(t) \pm \sqrt{\frac{1}{6 \pi}}  \left( \frac{3}{2} \log T - 3 \log a(t) \right)$ where it is significant in the final expression that the initial values may be replaced by the full coordinate time dependence; the invariants are constants of motion in the sense that they are independent of $t$. Also, $N(T)= N(t(T)) \frac{dt}{dT} =\frac{3}{2}  T^{1/2}$.

\item Dynamical variables may be gauge transformed through the use of the finite canonical generator $V_\xi(s,t) = \exp \left( s \{ - , G_\xi (t) \} \right)$. In particular, setting for $s =1$ the gauge transformed expansion factor $a_\xi^2$ equal to  $ t$, one can solve for the required dynamical-variable-dependent finite descriptor $\xi$. Employing this descriptor in the gauge transformation of the remaining variables we obtain the same invariant variables $\phi(t)$ and $N(t)$ as above.

\item Impose the gauge condition $t =  a^2(t)$. Preservation of this condition under time evolution leads to a new condition, $N= -\frac{3 \hbar}{4 \pi p_a}$. The Dirac-Bergmann procedure then yields the gauge-fixed Hamiltonian $H_{GF} = - \left( -\frac{\pi p_a^2}{3 a}  + \frac{ p_\phi^2}{2 a^3} \right) \frac{3}{4 \pi p_a} - p_N \frac{3 \hbar}{8 \pi a^2 p_a}$
with the equations of motion $\dot N = -\frac{3 \hbar}{8 \pi a^2 p_a}$, $\dot a = \frac{1}{2 a}$, $\dot p_a = -\frac{p_a}{2 a^2}$, 
$\dot \phi = - \frac{3 p_\phi}{4 \pi a^3 p_a}$, and $\dot p_\phi = 0$. The general solution is of course the same intrinsic time solution as above. 

\item Simply solve the constraints, taking $a(t) = t^{1/2}$ and $N(t) = 3 t^{1/2}/2$, leaving only $\phi$ as a dynamical variable. The corresponding reduced Lagrangian is $\hbar t \dot \phi^2/3$, with Hamiltonian $H(t) = \frac{3 p_\phi^2}{4 \hbar t}$. To recover the correct classical solutions one must in addition impose the condition that $p_\phi^2 = \frac{\hbar^2}{6 \pi}$.
\end{itemize}

\section{Generalized time-dependent Schr\"odinger equation}

Bojowald has shown that the expansion factor $a^2$ in the loop quantum gravitational approach has the discrete eigenvalues $t_k =  k/6$, where $k$ is a nonnegative integer.\cite{bojowald02}  We propose to employ the Hamiltonian obtained above to implement discrete time stepping. Thus we posit that 
\beq
|\psi(t_{k+1})>=\left(1-\frac{\imath}{\hbar}\Delta t \hat{H}(t_{k+1})\right)|\psi(t_k)>
=\left(1-\frac{9 \imath}{2 \hbar^2 (k+1)}\hat p_\phi^2 \right)|\psi(t_k)> ,\label{schro}
\eeq
We will work in a $\phi$ representation for which the operator $\hat p_\phi = \frac{\hbar}{\imath}\frac{\partial}{\partial \phi}$. The classical field $\phi$ can range from minus infinity to plus infinity. Our Hilbert space is thus 
$ {\bf L^2(\Re)}$. The minimum uncertainty state $ \psi( \phi, t_0) = (2 \pi \sigma^2)^{-1/4} \exp\left[ -\frac{(\phi - \phi_0)^2}{4 \sigma^2} + i  \frac{p_0 \phi}{\hbar} \right]$ can easily be shown to display the correct semi-classical behavior. We assume that the initial time $t_0 = \frac{6}{k+1}$, for large $k$. $\phi_0$ is the expectation value of $\hat \phi$ at time $t_0$, while $p_{0}$ is the expectation value of $\hat p_\phi$. One finds that
\beq
\int_{-\infty}^{\infty} |\psi(\phi, t_0+\triangle t)|^2 \phi d\phi  \approx \phi_0 + \frac{3 p_0}{2 \hbar t_0}  \triangle t,
\eeq
i.e., the expectation value satisfies the classical evolution equation. In addition, the expectation value of $\hat p_\phi$ is constant in time.  The classical correspondence limit requirement that $p_\phi^2 = \frac{\hbar^2}{6 \pi}$ can be imposed only as an expectation value. This supplementary condition would be removed in a more realistic massive scalar field model with a potential.

\section{Discussion and conclusions}

We have employed the reparameterization in time symmetry to argue that the imposition of an intrinsic time gauge condition produces reparameterization invariants. These invariants enjoy an evolution that can be modeled at the classical level, and promoted to a discrete quantum evolution. The lapse function itself undergoes a corresponding unique evolution; every choice of intrinsic time yields a fixed evolution in the lapse. Although in this model the lapse operator is merely a c-number function, in general it will be a non-trivial operator. Consequently it will generally undergo fluctuations. One might well question the legitimacy of this approach in which the intrinsic time, being itself a physical variable, does not itself seem to be subject to fluctuation. 

The only physical criterion employed in this construction is that the model yield the correct semiclassical limit. In this regard it is permissible to avoid the initial quantum singularity in the simple manner we have proposed; there is no time zero. The smallest time is $t_P/6$. The Bojowald difference equations that result from the imposition of the Hamiltonian constraint do not permit this choice.\cite{bojowald02} A detailed discussion of the relation between our construction, Bojowald's semiclassical limit, and the Wheeler-DeWitt equation will appear elsewhere.

\vfill

\end{document}